\documentclass[a4paper,12pt]{article}

\usepackage{a4}

\usepackage{latexsym, amssymb}

\usepackage{graphicx}
\usepackage{epsfig}

\newcommand{\lapprox}{%
\mathrel{%
\setbox0=\hbox{$<$}
\raise0.6ex\copy0\kern-\wd0
\lower0.65ex\hbox{$\sim$}
}}
\newcommand{\gapprox}{%
\mathrel{%
\setbox0=\hbox{$>$}
\raise0.6ex\copy0\kern-\wd0
\lower0.65ex\hbox{$\sim$}
}}

\newcommand{\be}{\begin{equation}}
\newcommand{\ee}{\end{equation}}
\newcommand{\bea}{\begin{eqnarray}}
\newcommand{\eea}{\end{eqnarray}}
\newcommand{\gev}{\mbox{GeV}}
\newcommand{\tev}{\mbox{TeV}}

\newcommand{\SM}{\mathrm{SM}}
\newcommand{\LHT}{\mathrm{LHT}}
\newcommand{\BR}{\mathrm{BR}}

\def\thefootnote{\fnsymbol{footnote}}

\begin{document}

\begin{titlepage}

\begin{flushright}
April 27, 2007 \\ 
HRI-P-06-11-003
\end{flushright}

\vspace*{0.2cm}
\begin{center}
{\Large {\bf Invisible Higgs boson decay in the \\[0.2cm]
 littlest Higgs model with T-parity}}\\[1cm] 
Raghavendra Srikanth Hundi\footnote{srikanth@mri.ernet.in}, Biswarup
Mukhopadhyaya\footnote{biswarup@mri.ernet.in} \\ 
and Andreas Nyf\/feler\footnote{nyf\/feler@hri.res.in} \\[0.4cm] 
{\it Harish-Chandra Research Institute \\
Chhatnag Road, Jhusi, 
Allahabad - 211 019, India}

\vspace*{1cm} 
\begin{abstract}

We study the invisible decay of the Higgs boson into a pair of stable, heavy
photons, $H \to A_H A_H$, in the littlest Higgs model with T-parity. For a
symmetry breaking scale of $f = 450~\mbox{GeV}$, the branching ratio
$H~\to~A_H A_H$ can be as high as $93\%$ for Higgs masses below
$150~\mbox{GeV}$. For $f = 500~\mbox{GeV}$, the invisible branching ratio is
about $75\%$ in the Higgs mass range $135 - 150~\mbox{GeV}$ and $10~(5.5)\%$
for $m_H = 200~(600)~\mbox{GeV}$. It drops to a few percent for $f$ larger
than $600~\mbox{GeV}$.  We have found regions in parameter space, allowed by
the electroweak precision data, with such low values of $f$ for
$115~\mbox{GeV} < m_H < 650~\mbox{GeV}$.

\end{abstract}

\end{center}

\pagestyle{plain}

\end{titlepage}

\setcounter{page}{1}


\renewcommand{\thefootnote}{\arabic{footnote}}
\setcounter{footnote}{0}

\section{Introduction}
\label{sec:intro}

One of the most crucial issues to be resolved in particle physics today is the
mechanism behind electroweak symmetry breaking. In the Standard Model (SM) the
symmetry is spontaneously broken by a fundamental Higgs boson which acquires a
vacuum expectation value (vev). Since the Higgs mass is not protected by any
symmetry, it turns out to be quadratically divergent as $\delta m_H^2 \sim
\Lambda^2$, $\Lambda$ being the cutoff scale for the SM.  This leads to the
well-known hierarchy or finetuning problem, whether
the SM is embedded in some grand unified theory (where $\Lambda \sim
10^{16}~\gev$) or not (where a cutoff at the Planck scale $M_P \sim
10^{19}~\gev$ must exist).

Then there is the so-called `little hierarchy problem'~\cite{LEP_paradox}. We
can view the SM as an effective field theory (EFT) with a cutoff $\Lambda$ and
parametrize new physics in terms of higher-dimensional operators which are
suppressed by inverse powers of the cutoff. Precision tests of the SM at low
energies and at LEP/SLC have not shown any significant deviations, which in
turn translates into a cutoff of about $\Lambda \sim 5-10~\mbox{TeV}$ which is
slightly more than an order of magnitude above the electroweak scale.

An attractive set of solutions to this little hierarchy problem are the little
Higgs models~\cite{LH_original, LH_reviews}. In these models, the Higgs boson
is a pseudo-Goldstone boson of a global symmetry which is spontaneously broken
at a scale $f$. This symmetry protects the Higgs mass from getting quadratic
divergences at one loop, even in the presence of gauge and Yukawa
interactions. The electroweak symmetry is broken via the Coleman-Weinberg
mechanism~\cite{Coleman_Weinberg} and the Higgs mass is generated radiatively,
which leads naturally to a light Higgs boson $m_H \sim (g^2 / 4\pi) f \approx
100~\mbox{GeV}$, if the scale $f \sim 1~\mbox{TeV}$. In contrast to
supersymmetric theories, here the new states at the TeV-scale which cancel the
quadratic divergences arising from the top quark, gauge boson and Higgs boson
loops, respectively, have the same spin as the corresponding SM particles. The
little Higgs model can then be interpreted as an EFT up to a new cutoff scale
of $\Lambda \sim 4 \pi f \sim 10~\mbox{TeV}$.

Among the different versions of this approach, the littlest Higgs
model~\cite{Littlest_Higgs} achieves the cancellation of quadratic divergences
with a minimal number of new degrees of freedom.  The global symmetry breaking
pattern is $SU(5) \to SO(5)$ and an $SU(2)_i \times U(1)_i, i=1,2,$ gauge
symmetry is imposed, that is broken down at the scale $f$ to the diagonal
subgroup $SU(2)_L \times U(1)_Y$, which is identified with the SM gauge
group. This leads to four heavy gauge bosons with masses $\sim f$ in addition
to the SM gauge fields. The SM Higgs doublet is part of an assortment of
pseudo-Goldstone bosons which result from the spontaneous breaking of the
global symmetry. The multiplet of Goldstone bosons contains a heavy $SU(2)$
triplet scalar as well.  Furthermore, a vectorlike heavy quark that can mix
with the top is postulated.  It turns out, however, that electroweak precision
data put very strong constraints on the littlest Higgs model.  Typically one
obtains the bound $f \gapprox 3-5~\mbox{TeV}$ in most of the natural parameter
space, unless specific choices are made for fermion representations or
hypercharges~\cite{LH_EW_tests}. Since $f$ effectively acts as a cutoff for
loops with SM particles, this reintroduces a little hierarchy between the
Higgs boson mass and the scale $f$.

The constraints from electroweak precision data can be bypassed by imposing a
discrete symmetry in the model, called T-parity~\cite{T_parity}.  In the
littlest Higgs model with T-parity (LHT)~\cite{LHT}, this discrete symmetry
exchanges the two pairs of gauge groups $SU(2)_i \times U(1)_i, i=1,2$,
forcing the corresponding gauge couplings to be equal $g_1 = g_2$ and
$g_1^\prime = g_2^\prime$. All SM particles, including the Higgs doublet, are
even under T-parity, whereas the four additional heavy gauge bosons and the
Higgs triplet are T-odd. The top quark has now two heavy fermionic partners,
$T_{+}$ (T-even) and $T_{-}$ (T-odd). For consistency of the model, one has to
introduce an additional heavy, T-odd vector-like fermion for each left-handed
SM quark and lepton field, see the original paper~\cite{LHT} and
Refs.~\cite{Hubisz_Meade,Hubisz_et_al} for details.

Two important phenomenological consequences of introducing T-parity are as
follows. First, if T-parity is exact, the lightest T-odd particle, typically
the heavy, neutral partner of the photon, $A_H$, is stable and can be a good
dark matter candidate~\cite{Hubisz_Meade,Asano_et_al,LH_Dark_Matter}.
Secondly, in the LHT there are no tree-level corrections to electroweak
precision observables and there is no dangerous Higgs triplet vev that
violates the custodial symmetry of the SM grossly. This relaxes the
constraints on the model from electroweak precision data and allows a
relatively small value of $f$ in certain regions of the parameter space.  Thus
T-parity opens up the possibility of seeing lighter new gauge bosons, scalars
and fermions (whose masses are related to the parameter $f$) than has in
general been thought possible in little Higgs scenarios.

In Ref.~\cite{Hubisz_et_al} it was shown that the LHT is compatible with
electroweak precision data, even for scales as low as $f \sim
500~\mbox{GeV}$. Although such a low value of $f$ may lead to additional
constraints on the model (such as a strong degeneracy of the heavy, T-odd
fermions, in order to avoid large flavor
violations~\cite{constraints_flavor_violations}) it is nevertheless an allowed
scenario.  Moreover, there are regions in parameter space where a Higgs boson
with a mass of $800~\mbox{GeV}$ is allowed, due to a cancellation in the
oblique $T$-parameter~\cite{Peskin_Takeuchi} between loops with a heavy Higgs
boson and the heavy T-even partner $T_{+}$ of the top quark.

This raises the interesting possibility that a heavy or even intermediate mass
Higgs boson could decay `invisibly' into a pair of stable, heavy photons
$A_H$. Such a possibility is facilitated by the fact that the state $A_H$ can
be quite light, with $M_{A_H} \approx g^\prime f / \sqrt{5} \approx 0.15
f$. Such T-odd particles can only be produced in pairs and lead to a signal of
large missing transverse energy~\cite{T_parity, Hubisz_Meade}, in a way
similar to the minimal supersymmetric standard model (MSSM). The decay $H \to
A_H A_H$ has already been mentioned briefly in
Refs.~\cite{Asano_et_al,Chen_Tobe_Yuan}. Ref.~\cite{Asano_et_al} obtained an
invisible branching ratio of about 5\% for $M_H \sim 170~\gev$, if the
condition is imposed that the heavy photon $A_H$ should constitute all of the
dark matter in the universe. Ref.~\cite{Chen_Tobe_Yuan} studied the production
and decay of the Higgs boson in the LHT at the LHC. However, the authors of
Ref.~\cite{Chen_Tobe_Yuan} only considered scales $f \gapprox 700~\gev$ where
the branching ratio is smaller than 1\% and therefore they did not take this
decay channel into account. Neither of these references did, however, consider
constraints from electroweak precision data for Higgs masses larger than
$115~\gev$.

The aim of this paper is to show that there are certain regions in the
parameter space of the LHT, compatible with electroweak constraints, where the
invisible decay $H \to A_H A_H$ can have a substantial branching ratio. This
ratio can be up to 95\% for an intermediate mass Higgs, and from 20\% down to
a few percents for a Higgs boson of mass $200~\gev$ or above. Such a high
invisible decay width is unlikely for the lightest neutral supersymmetric
Higgs, in the allowed range of its mass, at least in the minimal version of
the theory.  Therefore this invisible Higgs boson decay might help to
distinguish the LHT from the MSSM at present and future colliders. Of course,
this spectacular difference is noticeable only in the limited region of the
parameter space of the LHT, where $f \lapprox 600~\gev$.

The paper is organized as follows. In Section~\ref{sec:EW_Fits} we revisit
the electroweak precision tests applied to the LHT. We then calculate in
Section~\ref{sec:H_AHAH_Decay} the decay width $H \to A_H A_H$ and the
corresponding branching ratio for a range of values of $f$ allowed by the
precision data. We briefly discuss the consequences of our findings for Higgs
boson phenomenology at the Tevatron, the LHC and a future International Linear
Collider (ILC). We summarize and conclude in Section~\ref{sec:conclusions}.


\section{Electroweak precision tests revisited}
\label{sec:EW_Fits}

An analysis of the LHT in the light of the electroweak precision data has been
performed in Ref.~\cite{Hubisz_et_al}, based on the version of the model
proposed in Ref.~\cite{Hubisz_Meade}. In the following, we will closely follow
the notations and conventions of Ref.~\cite{Hubisz_et_al} and refer to
Refs.~\cite{Hubisz_Meade,Hubisz_et_al} and the original paper~\cite{LHT} for
an introduction into the LHT. Ref.~\cite{Hubisz_et_al} identified the regions
in the parameter space with a low symmetry breaking scale $f \sim
500~\gev$. However, such values of $f$ are obtained only for a very light
Higgs boson, $m_H = 113~\gev$. Since the T-odd heavy photon $A_H$ has a mass
of about $65~\gev$ for $f = 500~\gev$, the decay $H \to A_H A_H$ is
kinematically not possible. On the other hand, in Ref.~\cite{Hubisz_et_al}
some other regions in parameter space have been found where the Higgs boson is
very heavy ($m_H \sim 800~\gev$). Such a Higgs boson could easily decay into a
pair of heavy photons. In those regions, however, one also has $f \sim
1~\tev$. As we shall see in the next section, the decay width $\Gamma(H \to
A_H A_H)$ scales approximately like $1 / f^4$, so that the branching ratio is
very small for $f > 700~\gev$.

In our bid to identify the allowed regions giving high invisible branching
ratios, we have redone the electroweak fits, where we have found a small error
in the program used in Ref.~\cite{Hubisz_et_al}\footnote{We thank the
authors of Ref.~\cite{Hubisz_et_al} for providing us with the code of
their fitting program, which allowed us to track down the discrepancy.}. After
correcting for this error, we obtain slightly different allowed regions in
the parameter space. Although this does not affect the conclusions drawn in
\cite{Hubisz_et_al} in a qualitative manner, the allowed range of the Higgs
mass goes up by about $100~\gev$. 

Our $\chi^2$-fit is based on a parametrization~\cite{Burgess_et_al} of new
physics contributions to precision observables in terms of the oblique
parameters $S,T,U$~\cite{Peskin_Takeuchi} and a correction to the left-handed
$Zb\bar{b}$-vertex.  The expressions for these quantities in the LHT have been
calculated at one-loop in Ref.~\cite{Hubisz_et_al}. They depend on four
parameters: the symmetry breaking scale $f$, the mass of the Higgs boson
$m_H$, the ratio $R = \lambda_1 / \lambda_2$ of the two couplings
$\lambda_{1,2}$ that appear in the top-quark Lagrangian and a dimensionless
number $\delta_c$.  For $f \gg v_{\SM}$, the mass of the top quark is given by
$m_t = \lambda_1 v_{\SM} / \sqrt{R^2 + 1}$ and the mass of the heavy T-even
partner $T_{+}$ of the top is $m_{T_+} = \lambda_2 \sqrt{R^2 + 1} f$. Thus the
parameter $R$ is usually varied in the range $0 < R < 2$.  The quantity
$v_{\SM}$ is related to $v_{\LHT}$, the vev of the Higgs doublet in the LHT,
generated radiatively by the Coleman-Weinberg mechanism, as follows:
$v_{\SM} \equiv f \sqrt{1 - \cos (\sqrt{2} v_{\LHT} /
f)}$~\cite{Chen_Tobe_Yuan}.  Introducing $v_{\SM}$ allows one to express the
light gauge boson masses and the Fermi constant as in the SM at tree level:
$M_Z = \sqrt{g^2 + {g^\prime}^2} v_{\SM} / 2$, $M_W = g v_{\SM}/2$ and $G_F =
1 / (\sqrt{2} v_{\SM}^2)$, with $v_{\SM} \simeq 246~\gev$.

The parameter $\delta_c$ appears as the coefficient of a counterterm operator
that absorbs divergences in the $T$-parameter at one-loop in the LHT. These
divergences originate from the custodial $SU(2)$-violating tree-level mass
splitting of the T-odd heavy $W_H^3$ and $W_H^\pm$ gauge bosons. The absolute
value $|\delta_c|$ is assumed to be of order one and is varied here in the
range $-5 < \delta_c < 5$, in accordance with Ref.~\cite{Hubisz_et_al}.

In our fits we have considered the same 21 $Z$-peak and low-energy observables
as in Ref.~\cite{Hubisz_et_al} (see also Ref.~\cite{Burgess_et_al}).
However, we have used the new data from the PDG 2006~\cite{PDG_2006}.
Also, in accordance with the SM electroweak fits in Ref.~\cite{PDG_2006},
we have used the top quark mass $m_t = 172.7~\mbox{GeV}$. The reference value
of the Higgs boson mass in the oblique parameters $S,T,U$ has been taken as
the best SM fit value $m_{H,ref} = 89~\mbox{GeV}$.

\begin{figure}[!h]
\centerline{\epsfig{figure=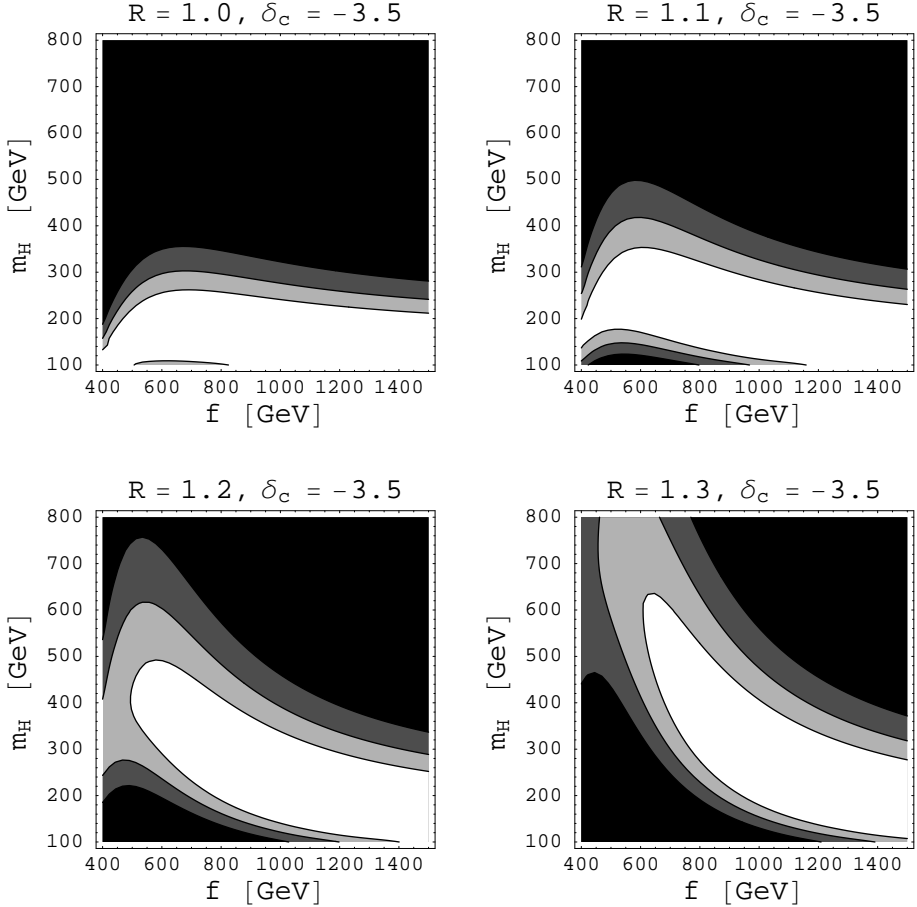,height=12cm,width=12cm}}
\caption{Exclusion contours in the plane of the Higgs mass $m_H$ and the
  symmetry breaking scale $f$ for $\delta_c = -3.5$ and four different values
  of $R$. From lightest to darkest, the contours correspond to the 95\%, 99\%,
  and 99.9\% confidence level exclusion.}
  \label{fig:EW_fits}
\end{figure}

Figure~\ref{fig:EW_fits} shows the constraints in the $f - m_H$ plane for
$\delta_c = -3.5$ and four different choices of the parameter $R$.  One
observes that there are allowed regions in parameter space where the symmetry
breaking scale $f$ is roughly between $400 - 700~\mbox{GeV}$ (or larger) and
$m_H$ is in the range $100-650~\mbox{GeV}$. Actually, in order to get allowed
regions with values of $m_H > 500~\mbox{GeV}$, it seems that the scale $f$ has
to be bigger than about $600~\mbox{GeV}$.  The choice of the parameters,
namely, $R \sim 1 - 1.3$, $\delta_c \sim -3.5$, is motivated by figures 5 and
6 in Ref.~\cite{Hubisz_et_al} where for $m_H = 113~\gev$ small allowed
values $f \lapprox 500~\gev$ had been found. The modified contour plots
obtained with our own program still have this feature. Varying $\delta_c$
between $-4$ and $-3$ leads to qualitatively similar conclusions. If $R \sim
1.1 - 1.3$, there are even allowed regions with low $f \sim 550~\gev$ for
positive values of $\delta_c \lapprox 3$, but only if the Higgs boson is not
too heavy, i.e.\ $m_H \lapprox 250~\gev$.

If $R \lapprox 0.5$ there are allowed regions with a rather small Higgs mass,
such as $m_H \sim 100 - 150~\gev$. This happens, however, 
only for $f > 700~\gev$, when the branching
ratio $H \to A_H A_H$ will be smaller than one percent. On the other hand, for
$R \gapprox 2$, our electroweak fits  show regions where very
large Higgs masses such as $m_H > 1~\tev$ are allowed. However, 
in those regions also $f > 1~\mbox{TeV}$, leading again 
to a very small branching ratio.
 

\section{The invisible decay $H \to A_H A_H$: results and implications} 
\label{sec:H_AHAH_Decay}

The mass (squared) of the heavy T-odd photon $A_H$ is given by
\be \label{massAH} 
M^2_{A_H} = {{g^\prime}^2 f^2 \over 5} - {{g^\prime}^2 v_{\SM}^2 \over 4}, 
\ee
neglecting higher powers of $v^2_{\SM}/f^2$. It can be checked that keeping
higher order terms in $M^2_{A_H}$ does not change our predictions
qualitatively. Since the heavy photon, as the lightest T-odd state, is stable,
there are no off-shell decays $H \to A_H^* A_H^*$ and the channel opens up
only for $m_H \geq 2 M_{A_H}$.  The interaction $H {A_H}_\mu A_H^\mu$ in the
LHT is described by the Feynman rule $(-i / 2) {g^\prime}^2 v_{\LHT}
g_{\mu\nu}$~\cite{Han_et_al, Hubisz_Meade}.  The decay width is then given by
\be \label{widthHtoAHAH} \Gamma(H \to A_H A_H) = { {g^\prime}^4 v^2_{\LHT}
\over 2048 \pi M_{A_H}^4} m_H^3 \beta_A \left( 4 - 4 a_A + 3 a_A^2 \right),
\ee
where $a_A = 1 - \beta_A^2 = 4 M_{A_H}^2 / m_H^2$. From
Eq.~(\ref{widthHtoAHAH}) we see that the partial width scales like $1/M_{A_H}^4
\sim 1/ f^4$. The quantity $v_{\LHT}$ is obtained from $v_{\SM}$ by inverting
the relation given earlier. 

As pointed out in Ref.~\cite{Chen_Tobe_Yuan}, the couplings of the Higgs boson
to the SM particles are subject to corrections in the LHT.  Using the
expressions for the couplings given in Ref.~\cite{Chen_Tobe_Yuan} one obtains
for the partial decay widths of the Higgs boson into SM gauge bosons ($V =
W, Z$) and into SM-fermions
\bea
\Gamma(H \to V V)_{\LHT} & = & \Gamma(H \to V V)_{\SM} \left( 1 - {1\over 4}
      {v_{\SM}^2 \over f^2} - {1 \over 32} {v_{\SM}^4 \over f^4} \right)^2, 
      \label{HVV} \\
\Gamma(H \to u \bar{u}, c \bar{c})_{\LHT} & = & \Gamma(H \to u \bar{u}, c
\bar{c})_{\SM} \left( 1 - {3\over 4} {v_{\SM}^2 \over f^2} - {5 \over 32}
    {v_{\SM}^4 \over f^4} \right)^2,  \\
\Gamma(H \to d \bar{d})_{\LHT} & = & \Gamma(H \to d \bar{d})_{\SM} \left( 1 - 
      {1\over 4} {v_{\SM}^2 \over f^2} + {7 \over 32} {v_{\SM}^4 \over 
    f^4} \right)^2 \label{Hdd}, \\
\Gamma(H \to t \bar{t})_{\LHT} & = & \Gamma(H \to t \bar{t})_{\SM} \left( 1 - {
  3 + 2 R^2 + 3 R^4 \over 4 (1 + R^2)^2} {v_{\SM}^2 \over f^2} \right)^2,
\label{Htt} 
\eea
up to higher corrections in $v^2_{\SM} / f^2$.  The expression for the decay
into down-type quarks in Eq.~(\ref{Hdd}) also applies to charged leptons.
This interaction is model-dependent, since there are different
ways~\cite{Chen_Tobe_Yuan} of introducing couplings of down-type quarks with
the Higgs boson that are consistent with the cancellation of quadratic
divergences and T-parity. Depending on the choice of the interaction terms,
the expression for the coupling in Eq.~(\ref{Hdd}) can numerically differ a
little or considerably from the SM value. Since we are interested here in the
effects of the invisible decay $H \to A_H A_H$, we have taken in
Eq.~(\ref{Hdd}) the couplings of ``case A'' from Ref.~\cite{Chen_Tobe_Yuan},
where the decay width in the LHT is not much suppressed compared to the
SM. For the evaluation of the decay mode $H \to t \bar{t}$ we will take $R =
1$ in Eq.~(\ref{Htt}).

The loop-induced decay $H \to g g$ gets further modified, since the additional
T-even and T-odd fermions in the LHT can also run in the loop. It was
shown in Ref.~\cite{Chen_Tobe_Yuan} that for fermion masses much heavier
than the Higgs boson, the full one-loop result can be well described by the
approximate formula
\be \label{Hgg} 
\Gamma(H \to g g)_{\LHT} =  \Gamma(H \to g g)_{\SM} \left( 1 - 3 {v_{\SM}^2
  \over f^2} \right). 
\ee

In the decays $H \to \gamma\gamma$ and $H \to Z \gamma$, the $W$-boson loop
dominates over the contribution from the top quark~\cite{Spira}. For the decay
into two photons this is still true if the contributions from the heavy
fermions in the LHT are taken into account~\cite{Chen_Tobe_Yuan}. We therefore
simply rescale both decay widths with the same factor as in Eq.~(\ref{HVV}). 

We have used the program HDECAY~\cite{HDECAY} to calculate the partial widths
of the Higgs boson in the SM in the various channels. The program includes all
relevant higher order QCD and electroweak corrections.  The corresponding
decay widths in the LHT have then been obtained in a simplified (and
approximate) way by multiplying the SM results with the correction factors
from Eqs.~(\ref{HVV})--(\ref{Hgg}). Adding the new invisible decay mode
$H \to A_H A_H$ from Eq.~(\ref{widthHtoAHAH}) leads to the total width of
the Higgs boson in the LHT, which, as we shall see later, can change the total
width by as much as an order of magnitude. We do not expect the
conclusions we draw in the following to change qualitatively, if one were to
include all radiative corrections within the LHT, e.g.\ not simply taking the
tree-level expression~(\ref{widthHtoAHAH}) for the mode $H \to A_H A_H$ and
rescaling the partial widths in the SM, Eqs.~(\ref{HVV})--(\ref{Hgg}).

\begin{figure}[!t]

\epsfig{figure=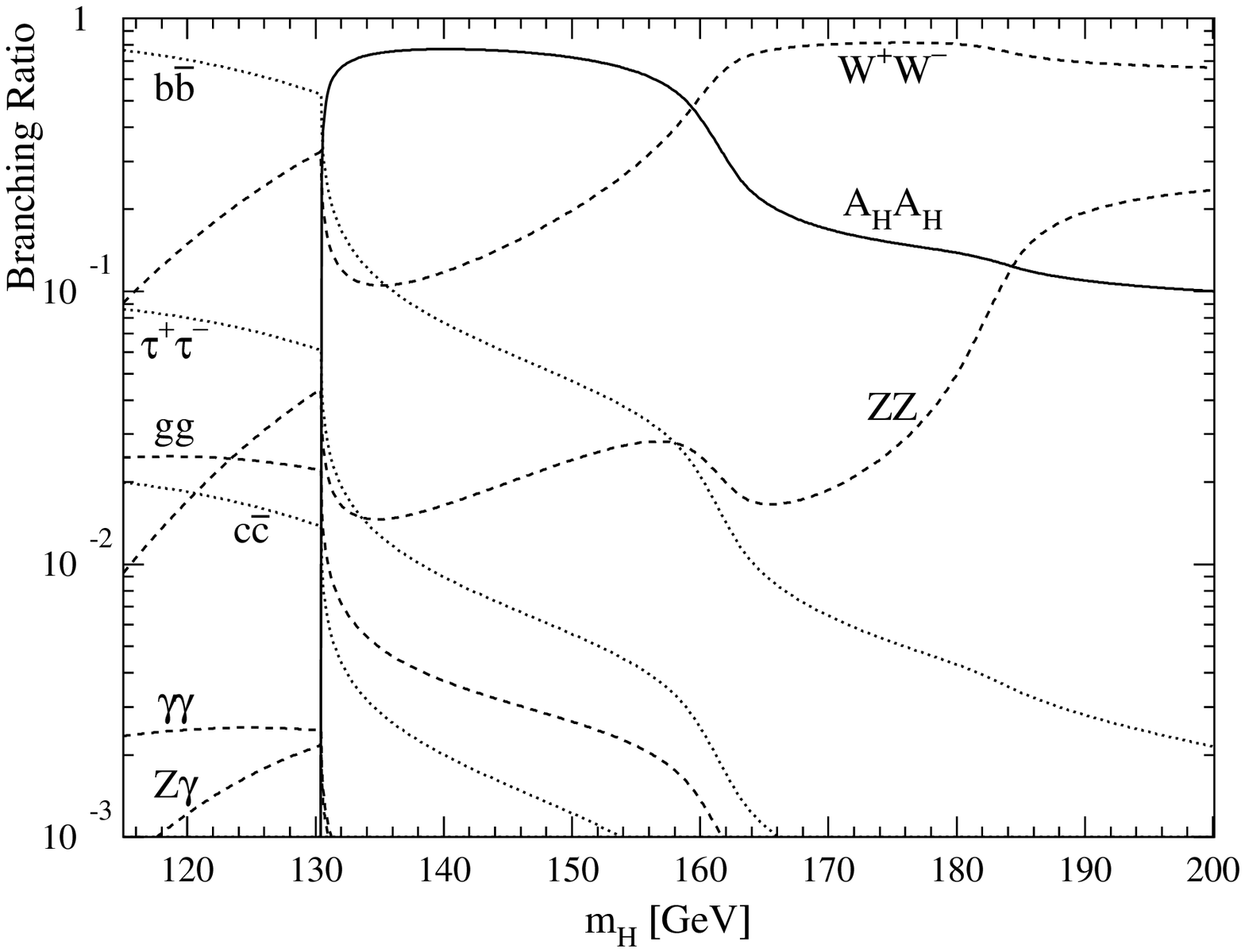,height=7.5cm,width=7.5cm}
\epsfig{figure=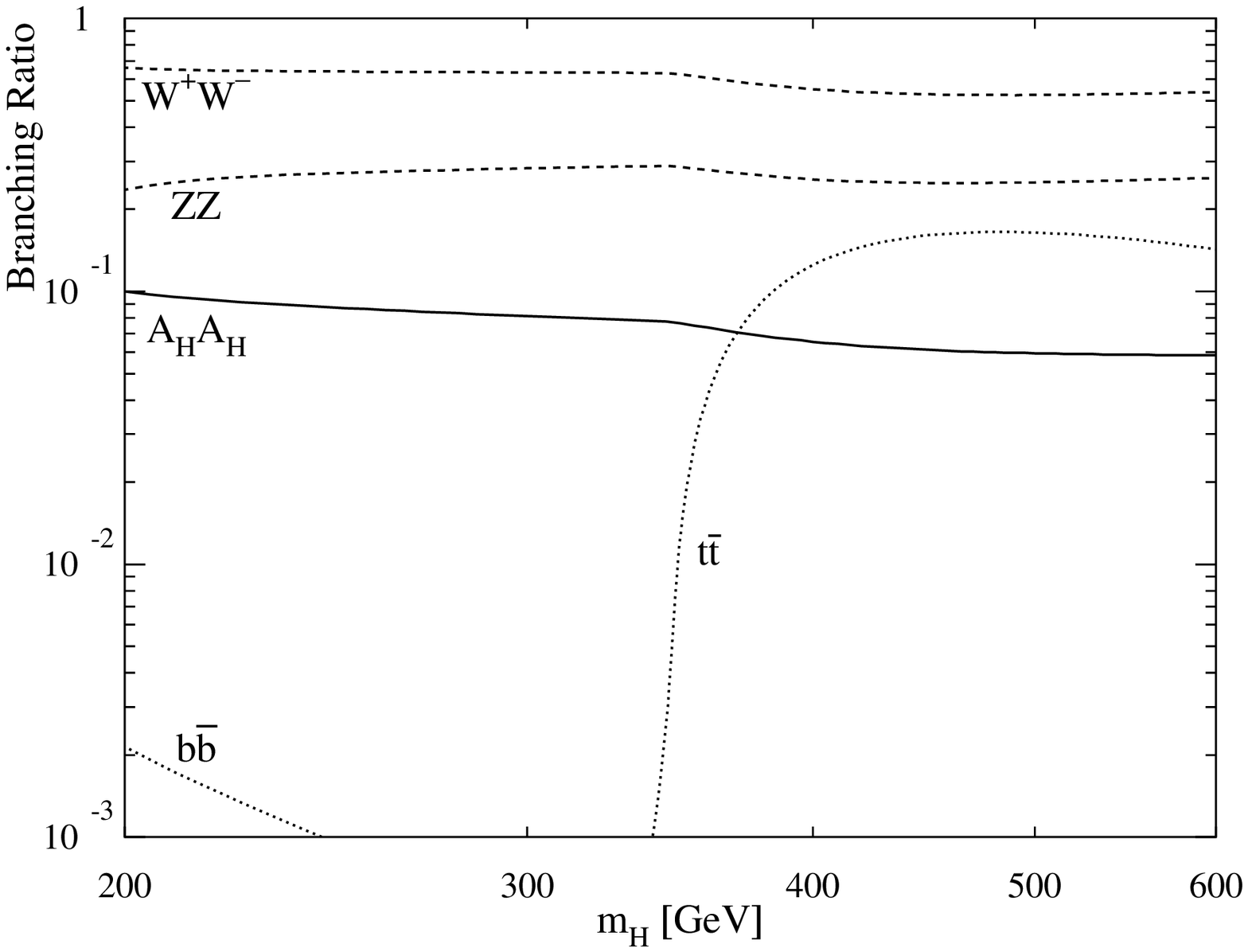,height=7.5cm,width=7.5cm}

\caption{Branching ratios in the littlest Higgs model with T-parity for Higgs
  masses below $200~\gev$ (left panel) and above  $200~\gev$ (right panel)
  for a symmetry breaking scale $f = 500~\gev$.} 
  \label{fig:BR_LHT}
\end{figure}

In Fig.~\ref{fig:BR_LHT} we have plotted for $f=500~\mbox{GeV}$ all branching
ratios of the Higgs boson in the LHT that are larger than $10^{-3}$ in the
mass range $115~\gev < m_H < 600~\gev$. One observes, that as soon as the
decay $H \to A_H A_H$ is kinematically allowed, $m_H \geq 2 M_{A_H} =
130~\mbox{GeV}$, we get a huge invisible $\BR(H \to A_H A_H)$ of about $75\%$
in the Higgs mass range $135 - 150~\gev$. The reason is that the Higgs boson
couples to the heavy photons $A_H$ with electroweak strength $g^\prime$ which
is much larger than the Yukawa coupling to the bottom quarks. The decay width
is also larger than the off-shell (three or four-body) decay $H \to W^{(*)}
W^{*}$, unless that decay starts to grow around $m_H = 2 M_W$. At $m_H =
159~\gev$ we have $\BR(H \to A_H A_H) \approx \BR(H \to WW) = 47\%$. At $m_H =
200~(600)~\mbox{GeV}$ the invisible decay BR is still about $10~(5.5)\%$. We
would like to stress that there are regions in parameter space where values of
$115~\gev < m_H < 600~\mbox{GeV}$ and $f = 500~\mbox{GeV}$ are allowed by the
electroweak data, see Fig.~\ref{fig:EW_fits}.

Below the threshold of $130~\gev$, the same decay channels are open as in the
SM, however, $H \to gg$ is highly suppressed in the LHT, see
Ref.~\cite{Chen_Tobe_Yuan}. On the other hand, the branching ratio for $H \to
\gamma\gamma$ for $m_H = 115~\gev$ is about $2.3 \times 10^{-3}$, i.e.\ very
close to the one in the SM. Since we have taken the fermion couplings in
Eq.~(\ref{Hdd}) from the ``case A'' proposed in Ref.~\cite{Chen_Tobe_Yuan},
which differ not much from their SM values, there is no large enhancement of
the $H\to \gamma\gamma$ mode as observed in that reference for the ``case
B''. As soon as the decay $H \to A_H A_H$ is possible, all other branching
ratios drop down considerably.

Note that we have not taken into account the off shell-decays $H \to W_H^{*}
W_H^{*}$ and $H \to Z_H^{*} Z_H^{*}$ for Higgs masses larger than $M_{W_H} =
M_{Z_H} = 317~\gev$ for $f = 500~\gev$. We expect the corresponding branching
ratios to be small below the two-particle threshold.

\begin{figure}[!t]

\epsfig{figure=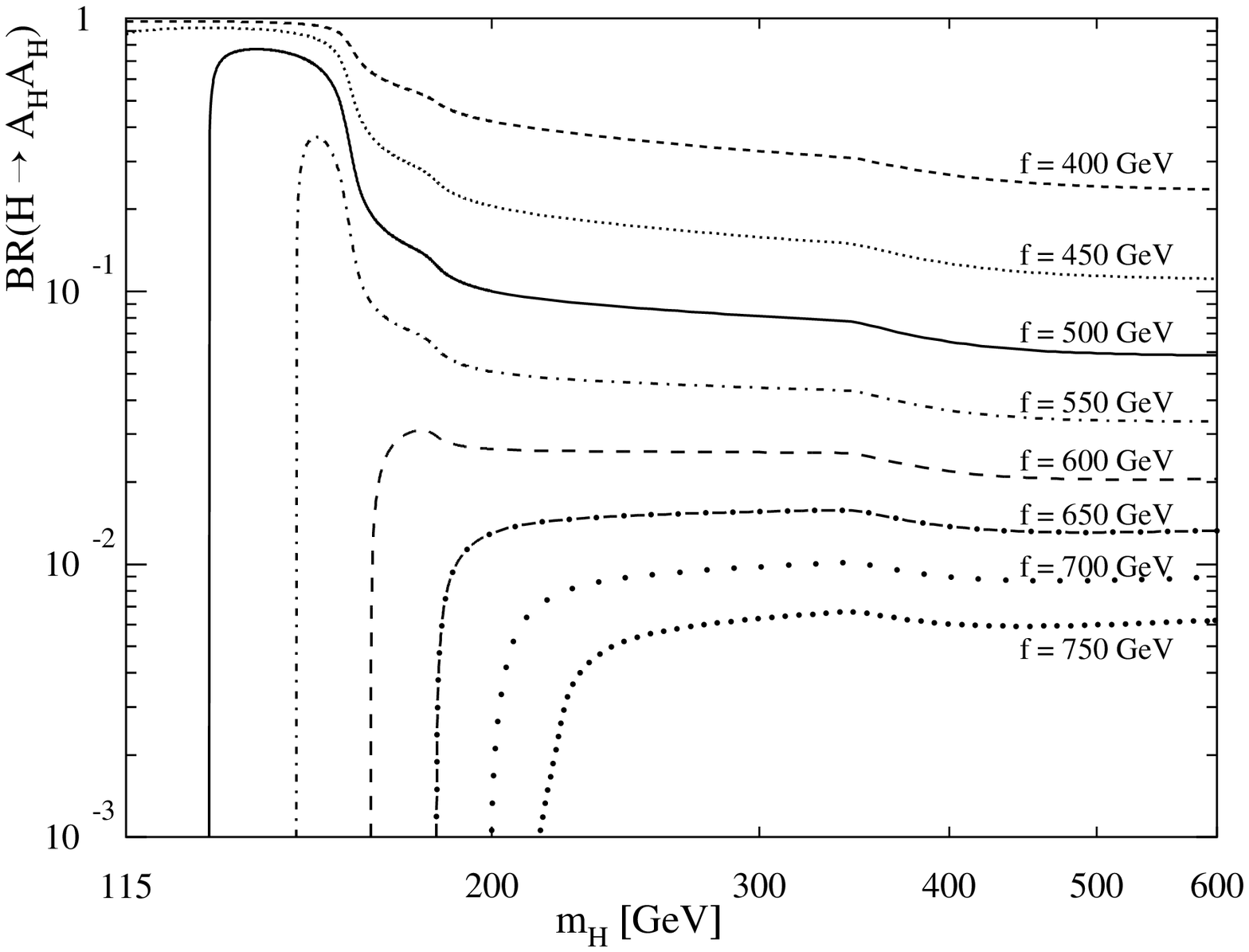,height=7.5cm,width=7.5cm}
\epsfig{figure=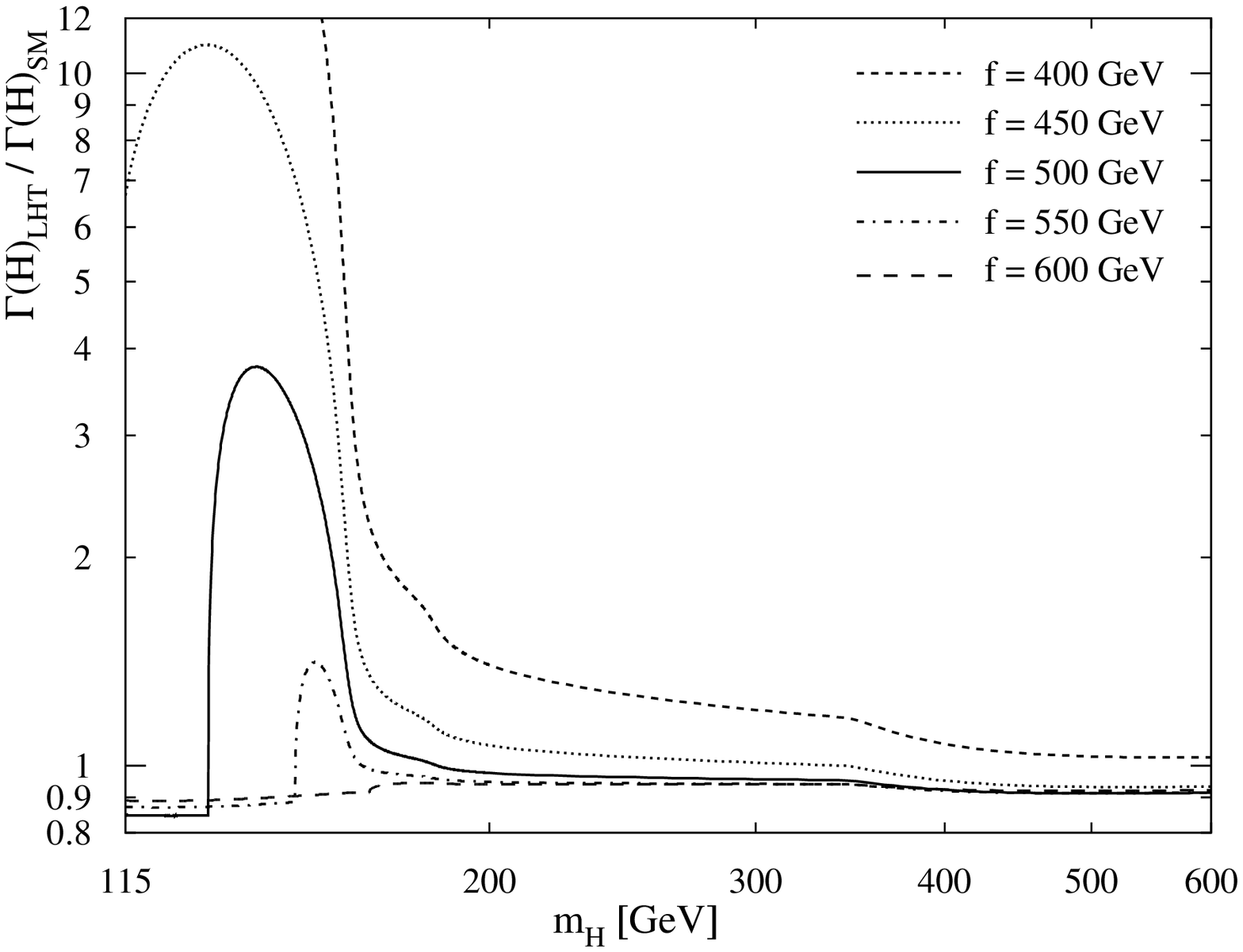,height=7.5cm,width=7.5cm}

\hspace*{3.35cm}{\small (a)}\hspace*{7.1cm}{\small (b)} 

\caption{(a) Branching ratio for the invisible decay $H \to A_H A_H$ in the
  littlest Higgs model with T-parity for several values of the symmetry
  breaking scale~$f$. (b) Ratio of the total decay width of the Higgs boson in
  the LHT, $\Gamma(H)_{\LHT}$, to the total decay width in the SM,
  $\Gamma(H)_{\SM}$, for different values of $f$. The curve for $f = 400~\gev$
  peaks at a value of about 35 for $m_H \approx 126~\gev$.}
\label{fig:BR_AHAH_Gamma_ratio}

\end{figure}

In Fig.~\ref{fig:BR_AHAH_Gamma_ratio}(a) we show the invisible branching
ratio $H \to A_H A_H$ as a function of the Higgs mass for different
values of $f$ in the range $400 - 750~\mbox{GeV}$. For $f =
400~(450)~\mbox{GeV}$ the branching ratio can be as large as 98~(93)\% for
Higgs masses below about $150~\mbox{GeV}$. Although values of $f =
400~\mbox{GeV}$ are allowed by the precision data (see
Fig.~\ref{fig:EW_fits}), the calculation cannot be completely trusted
there. For such low values of $f$, higher derivative terms in the low-energy
expansion in the LHT model, generated at the scale $\Lambda \sim 4\pi f$,
should be taken into account.  It may be noted that values of $f \lapprox
400~\mbox{GeV}$ have been discarded in Ref.~\cite{Hubisz_et_al} for this
reason. In this sense $f \simeq 400~\gev$ marks the lower end of the parameter
space where the underlying framework is reliable. 

For $f = 600~\gev$, the invisible branching ratio is $2-3\%$ for $m_H \gapprox
169~\gev$, whereas it drops below 1\% for $f \geq 700~\mbox{GeV}$. Since the
$A_H A_H$-threshold for $f= 600~\gev$ is about $167~\mbox{GeV}$, i.e.\ above
the $WW$-threshold, the on-shell decays into $WW$ and later into $ZZ$
overwhelm the invisible decay $H\to A_H A_H$.  Note that for $f =
650~\mbox{GeV}$ we have $M_{A_H} \approx M_Z$ and therefore $\Gamma(H \to A_H
A_H) \approx \left( {g^\prime}^2 c_W^2 / g^2 \right)^2 \Gamma(H \to Z Z)_{\SM}
\approx 0.05 \, \Gamma(H \to Z Z)_{\SM}$, i.e.\ the BR is $1.3\%$ at $m_H =
200~\gev$.

Figure~\ref{fig:BR_AHAH_Gamma_ratio}(b) shows the ratio of the total decay
width of the Higgs boson in the LHT, $\Gamma(H)_{\LHT}$, to the total width in
the SM, $\Gamma(H)_{\SM}$, as a function of the Higgs mass for a subset of
values of $f$ used in Fig.~\ref{fig:BR_AHAH_Gamma_ratio}(a). In contrast to
earlier studies~\cite{Higgs_decays_LH,Chen_Tobe_Yuan} which always observed a
reduction of the total decay width of the Higgs boson in the littlest Higgs
model and in the LHT compared to the SM, we get a potentially huge enhancement
of the decay width for values $f \leq 550~\mbox{GeV}$.  For $f = (400, 450,
500, 550)~\gev$, the maximal enhancement factors of $(34.8, 11.0, 3.77, 1.41)$
that can be seen in Fig.~\ref{fig:BR_AHAH_Gamma_ratio}(b) correspond to
$\Gamma(H)_{\LHT} = (140, 51, 30, 33)~\mbox{MeV}$ at $m_H = (125.8, 130.2,
140.5, 153.4)~\gev$. Note, however, that the width of the Higgs boson in the
SM is very small for Higgs masses below the $WW$-threshold.  Only for $f \geq
600~\gev$ we obtain a reduction of the total width for the whole range of
Higgs masses $115~\gev < m_H < 600~\gev$. The ratio $\Gamma(H)_{\LHT} /
\Gamma(H)_{\SM}$ varies between $0.89$ and $0.95$ for values of $f = 600 -
750~\gev$.

A substantial branching ratio into the invisible channel not only makes the
Higgs boson a rather interesting object but also helps in associating it with
some specific types of non-standard physics.  For example, in supersymmetric
theories, the lightest neutral scalar can in principle decay into two lightest
neutralinos, making it invisible.  However, the branching ratio of such a
decay is usually not very high, and is rather restricted in the regions of the
parameter space allowed by LEP data, at least in those versions of the theory
not too far from the minimal model. In our case, however, the invisible
branching ratio can not only be appreciable but also may correspond to a Higgs
boson that is heavier than what is allowed in a minimal supersymmetric
framework. Thus this region of the parameter space may provide a test to
distinguish the LHT from a supersymmetric scenario.

In the context of LEP, invisible Higgs boson decays have been studied earlier,
and it has been concluded that a Higgs boson is identifiable in the channel
$e^{+}e^{-} \to ZH$, with the $H$ decaying invisibly, and the $Z$ decaying
into $q{\bar q}$ (acoplanar jets) or a lepton pair. The current
limit~\cite{LEP_limit} of $m_H > 114.4~\gev$ at $95\%$ confidence level for an
invisible Higgs boson assumes SM production rates and a 100\% invisible
branching ratio. The small modification of the $HZZ$ coupling in the LHT will
not change this limit significantly. A hadron collider, on the other hand,
will find it a more challenging task to unravel a Higgs boson that has a
dominantly invisible decay mode. Channels such as associated production with
gauge bosons, $p p \to WH, ZH$~\cite{pp_WH_ZH} or top quarks $p p \to t\bar{t}
H$~\cite{pp_ttH} have been suggested in this context.  Another possibility is
to consider Higgs production by weak boson fusion (WBF)~\cite{VBF} and look
for final states with two energetic forward jets with a large rapidity gap and
a large amount of missing energy in the gap. The efficacy of this search
strategy has been discussed in the context of not only an invisible Higgs
boson but also other types of new physics signals leading to invisible final
states~\cite{invisible_final_states}. Combining WBF and associated production
with $Z$ might even allow the discovery at the Tevatron or in the early phase
of LHC~\cite{Davoudiasl_Han_Logan}. The biggest potential, however, lies in a
linear collider~\cite{invisible_Higgs_ILC}, where an invisible Higgs boson is
identifiable in the form of a peak recoiling against a $Z$-peak marked by a
fermion-antifermion pair. The branching ratio of the Higgs boson decaying into
the invisible channel can also be measured in this way, thus making it
possible to probe the spectrum and new interactions of an LHT scenario (or any
other theory that has similar effects).

 
\section{Summary and conclusions}
\label{sec:conclusions}

In this paper we have studied within the littlest Higgs model with T-parity,
the decay of the Higgs boson into a pair of the heavy, T-odd partners of the
photon, $H \to A_H A_H$. The neutral and stable heavy photons $A_H$ will not
leave any traces in the detector, thereby leading to an invisible decay mode
of the Higgs boson with a missing energy signature. For Higgs masses between
$115$ and $650~\gev$ we have found regions in the parameter space, compatible
with the latest electroweak precision data at 95\% confidence level, where the
symmetry breaking scale in the LHT can be quite low $f \sim
400-700~\gev$. Such values of $f$ then lead to a substantial invisible
branching ratio. For $f = 450~\gev$, the $\BR(H \to A_H A_H)$ can be as high
as $93\%$ for Higgs masses below $150~\gev$. The invisible decay takes over as
soon as it is kinematically allowed. This is due to the fact that the
interaction of the Higgs boson with the heavy photons $A_H$ is governed by the
gauge coupling $g^\prime$ which is much larger than the $b$-quark Yukawa
coupling. Furthermore, the decay width $H \to A_H A_H$ scales like $1 /
M_{A_H}^4 \sim 1/f^4$. Because of phase space, it also dominates over the
off-shell mode $H \to W^{(*)} W^{*}$, almost up to the $2 M_W$ threshold. For
$f = 500~\gev$, the $\BR(H \to A_H A_H)$ is about $75\%$ in the Higgs mass
range $135 - 150~\gev$ and $10~(5.5)\%$ for $m_H = 200~(600)~\gev$. For $f =
600~\gev$, the invisible branching ratio is $2-3\%$ above $m_H = 169~\gev$ and
it drops below one percent for $f \geq 700~\gev$. For $f = 450, 500$ and
$550~\gev$, respectively, the total decay width of the Higgs boson in the LHT
can be enhanced by a factor of $11, 3.8$ and $1.4$ (at $m_H = 130, 141$ and
$153~\gev$) compared with the SM.

Our analysis did not take into account the requirement that the heavy photon
$A_H$ has to constitute all of the dark matter in the universe. If such a
condition is imposed then one may have to restrict oneself to values of $f
\gapprox 580~\gev$, which corresponds with Eq.~(\ref{massAH}) to $M_{A_H}
\gapprox 80~\gev$, see Refs.~\cite{Hubisz_Meade, Asano_et_al,
LH_Dark_Matter}. The invisible branching ratio $H \to A_H A_H$ including the
bounds from dark matter has been evaluated in Ref.~\cite{Asano_et_al},
however, without considering the constraints from electroweak precision data.

Such an invisible decay mode of the Higgs boson has been studied earlier in
the context of several models of new physics beyond the Standard Model (e.g.\
MSSM, extra dimensions, \ldots). Although general procedures have been
developed to detect such a signal at $e^+ e^-$ machines (LEP, ILC) and hadron
colliders (Tevatron, LHC), this specific decay $H \to A_H A_H$ in the
littlest Higgs model with T-parity certainly deserves detailed further studies
which might help to determine the parameters of the LHT and to distinguish it
from other models.


\section*{Acknowledgements} 

We would like to thank Santosh Kumar Rai for discussions and V.\ Ravindran for
carefully reading the manuscript. Furthermore, we are grateful to J.\ Hubisz,
P.\ Meade, A.\ Noble and M.\ Perelstein for their kind cooperation in order to
find the differences in our programs used for the electroweak fits. This work
was partially supported by the Department of Atomic Energy, Government of
India, under the X-th five-years plan.


\end{document}